\documentclass[12pt]{article}

\usepackage{amsfonts,amssymb,amsmath,amsthm,amscd}
\usepackage{color}
\usepackage{hyperref}
\usepackage{verbatim}

\newtheorem{Theorem}{Theorem}
\newtheorem{Definition}{Definition}

\newtheorem{Lemma}{Lemma}

\newcommand{\calF}{\mbox{${\mathcal F}$}}
\newcommand{\calU}{\mbox{${\mathcal U}$}}
\newcommand{\calB}{\mbox{${\mathcal B}$}}
\newcommand{\calM}{\mbox{${\mathcal M}$}}
\newcommand{\calK}{\mbox{${\mathcal K}$}}
\newcommand{\calO}{\mbox{${\mathcal O}$}}

\newcommand{\DerH}{\mbox{$\mathrm{Der}_H$}}

\newcommand{\Der}{\mbox{$\mathrm{Der}$}}
\newcommand {\calA}{\mbox{${\mathcal A}$}}

\newcommand {\calH}{\mbox{${\mathcal H}$}}
\newcommand {\calZ}{\mbox{${\mathcal Z}$}}

\newcommand {\cM}{\mbox{${\mathcal M}$}}

\newcommand{\C}{\mathbb{C}}
\newcommand{\N}{\mathbb{N}}
\newcommand{\R}{\mathbb{R}}

\newcounter{mnotecount}[section]
\renewcommand{\themnotecount}{\thesection.\arabic{mnotecount}}
\newcommand{\mnote}[1]
{\protect{\stepcounter{mnotecount}}$^{\mbox{\footnotesize
$
\bullet$\themnotecount}}$ \marginpar{
\raggedright\tiny\em
$\!\!\!\!\!\!\,\bullet$\themnotecount: #1} }

\begin{document}

\title{A Sheaf of von Neumann Algebras and Its Geometry}
\author{Micha{\l} Eckstein \\
\url{michal.eckstein@uj.edu.pl} \\
Faculty of Mathematics and Computer Science, \\ Jagellonian University, \\
ul. {\L}ojasiewicza 6, 30-348 Krak\'ow, Poland \\
and Copernicus Center for Interdisciplinary Studies \\
ul. S{\l}awkowska 17, 31-016 Krak\'ow, Poland; \\
\and Michael Heller, \\
\url{mheller@wsd.tarnow.pl} \\
Copernicus Center for Interdisciplinary Studies \\
ul. S{\l}awkowska 17, 31-016 Krak\'ow, Poland \\
and Vatican Observatory, V-00120 Vatican City State;\\ 
\and Leszek Pysiak, \\
\url{lpysiak@mini.pw.edu.pl}\\
Technical University of Warsaw, \\
Plac Politechniki 1, 00-661 Warszawa, Poland \\
and Copernicus Center for Interdisciplinary Studies \\
ul. S{\l}awkowska 17, 31-016 Krak\'ow, Poland; \\
and Wies{\l}aw Sasin, \\
Technical University of
Warsaw,\\ Plac Politechniki 1, 00-661 Warszawa, Poland \\
and Copernicus Center for Interdisciplinary Studies \\
ul. S{\l}awkowska 17, 31-016 Krak\'ow, Poland \\
}

\date{\today}
\maketitle

\begin{abstract}
It is shown that the differential geometry of space-time, can be expressed in terms of the algebra of operators on a bundle of Hilbert spaces. The price for this is that the algebra of smooth functions on space-time $M$ has to be made noncommutative. The generalized differential geometry of space-time is constructed in terms of the algebra \calA \ (and its derivations) on a transfrmation groupoid. Regular representation $\pi $ of \calA \ in the algebra of bounded operators on a bundle of Hilbert spaces  leads to the algebra $\pi (\calA) = \calM_0$ which can be completed to the von Neumann algebra \calM . The representation $\pi $ establishes the isomorphism between \calA \ and $\calM_0$ which, in turn, implies the isomorphism between moduli of their derivations. In this way, geometry naturally transfers to the algebra $\calM_0$ and its derivations. Although geometry, as defined in terms of $\calM_0$, is formally isomorphic to that defined in terms of \calA , it exhibits a strong probabilistic flavour. However, the geometry of $\calM_0$ does not prolong to \calM . This is clearly a serious stumbling block to fully unify mathematical tools of general relativity and quantum theory.
\end{abstract}

\section{Introduction}
Mathematical tool of general relativity is differential geometry of space-time manifold $M$, which can be formulated either in terms of maps and atlases on $M$ or, equivalently, in terms of the algebra $C^{\infty }(M)$ of smooth functions on $M$ \cite{Geroch72}, whereas the standard tool of quantum mechanics and quantum field theories are operator algebras (usually $C^*$-algebras or von Neumann algebras). These different mathematical structures of general relativity and quantum physics are one of major stumbling blocks preventing their unification. This is why any study that sheds light onto mutual relations between differential geometry and operator algebras seems to be of high interest. In the present paper we show that the standard tool of general relativity, i.e., the differential geometry of space-time, can be expressed in terms of the algebra of operators on a bundle of Hilbert spaces which is very close to the standard tools of quantum physics. The price for this rapprochement is that the algebra $C^{\infty }(M)$ of smooth functions on space-time $M$ has to be made noncommutative. This is done by constructing a groupoid $\Gamma = E \times G$, where $E$ is the frame bundle (with $G$ as its structural group) over space-time $M$, and defining the algebra \calA \ of smooth, compactly supported functions on $\Gamma $ with convolution as multiplication. The generalized differential geometry of space-time is constructed in terms of this algebra and its derivations. This is described in section 2. Regular representation $\pi $ of \calA \ in the algebra of bounded operators on a bundle of Hilbert spaces  leads to the algebra $\pi (\calA) = \calM_0$ which can be completed to the von Neumann algebra \calM . In section 3, we demonstrate that elements of the algebra $\calM_0$ are random operators in the sense of Connes \cite{Connes}. 

In section 4, we show that the algebra $\calM_0(U)$, with $U$ open in $M$, is of the form $\calM_0(U) = C^{\infty }_b(U) \otimes \calK (C^{\infty }_c(G \times G))$, where $C^{\infty }_b(U)$ is an algebra of smooth bounded functions on $U$, $\calK (C^{\infty }_c(G \times G))$ is an algebra of integral operators with kernel belonging to $C^{\infty }_c(G \times G)$, and we use the fact that $\calM(U) = L^{\infty }(U) \bar{\otimes } \calB(L^2(G))$ (demonstrated in Appendix C) to study geometry of the von Neumann algebra \calM . To this end we employ the technique of sheaves (some elements of it are recalled in Appendix A). The representation $\pi $ establishes the isomorphism between \calA \ and $\calM_0$ which, in turn, implies the isomorphism between moduli of their derivations. Since the generalized geometry of space-time is defined in terms of the algebra \calA\ and its derivations, it naturally transfers to the algebra $\calM_0$. Geometry, as defined in terms of $\calM_0$  and its derivations, is formally isomorphic to that defined in terms of \calA \ and its derivations, but it exhibits a new interpretative element, namely it has a strong probabilistic flavour (algebra $\calM_0$ consists of random operators).

Can this geometry be extended from $\calM_0$ to the full von Neumann algebra \calM ? Outer derivations of $\calM_0$ do not prolong to \calM , and since the standard differential geometry (connection, curvature, etc.) is defined with the help of outer derivations, this geometry breaks down at \calM . Only inner derivations survive the completion to \calM \ (as is well known, von Neumann algebras admit only inner derivations). The standard space-time geometry being commutative, has only vanishing inner derivations and, consequently, it has no contact with $\calM - \calM_0$. This is clearly a serious stumbling block to fully unify mathematical tools of general relativity and quantum theory, at least according to lines proposed in the present work.

In section 5, we illustrate our results with a simple example of a matrix algebra on a groupoid $\Gamma_U = U \times G \times G$ where $G$ is a finite group.

\section{Geometry of a Groupoid Algebra}
Let $M$ be a smooth manifold, and $\pi_M: E \rightarrow M$ the frame bundle over $M$ with the structure group $G$. Since $G$ acts on $E$ (to the right), $E \times G \rightarrow E$, we can equip $\Gamma = E \times G$ with the groupoid structure. The groupoid obtained in this way is called transformation groupoid \cite[p. 90]{Weinstein}. Let further $\calA = C_c^{\infty }(\Gamma , \C)$ be an algebra of compactly supported, smooth, complex valued functions on the groupoid $\Gamma $ with the convolution
\[
(f * g)(\gamma ) = \int_{\Gamma^{r(\gamma )}}f(\gamma_1)g(\gamma_1^{-1} \circ \gamma )d\gamma_1,
\]
$f, g \in \calA $, as multiplication. Here $\gamma = (p,g) \in \Gamma $, $r(p,g) = pg$ (and analogously, $d(p,g) = p$), $\Gamma^q = \{\gamma \in \Gamma :r(\gamma ) = q, q \in E\}$ (and analogously $\Gamma_q = \{\gamma \in \Gamma :d(\gamma ) = q, q \in E\}$) and $d\gamma_1$ denotes the Haar system on the groupoid $\Gamma $. The symbol $\Gamma^{r(\gamma )}$ denotes elements  of $\Gamma $ that end at $r(\gamma )$.  This algebra is, in general, noncommutative, and its center is null, ${\cal Z}({\cal A})=\{0\}$ (see Appendix B), but ${\cal A}$ is a module over $Z=\pi_M^{*}(C^{\infty}(M))$. Functions of $Z$, which in general are not compactly supported, act on ${\cal A}$, $\alpha :Z\times {\cal A} \rightarrow {\cal A}$, in the following way
\[\alpha (f, a)(p,q)=f(p)a(p,g),\]
$f\in Z,\,a\in {\cal A}$. In \cite{FullJMP} and \cite{Inner}, we have constructed a derivation-based geometry in terms of the algebra \calA \ and its derivations in close analogy to the differential geometry of the manifold $M$ when it is constructed in terms of the algebra $C^{\infty }(M)$ and its derivations.

We now briefly summarize the construction of the the differential geometry in terms of the algebra \calA \ and its derivations; for details the reader should refer to \cite{FullJMP,Inner}.

In the $Z$-module of derivations of the algebra \calA \ we can distinguish three types of derivations:

(i) Horizontal derivations. With the help of the connection in the frame bundle $\pi _{M}:E\rightarrow M$ we lift a vector field $X$ on $M$ to $E$, i.e., $\bar{X }(p)=\sigma (X(x)),\,x = \pi _{M}(p)\in M$, where $\sigma $ is a chosen lifting
homomorphism. This vector field is right invariant on $E$. After being lifted further to $\Gamma $
\[
\bar{\bar{X}}(p,g)=(\iota _{g})_{\ast p}\bar{X}(p),
\]
it becomes a left invariant derivation of the algebra $\mathcal{A}$. We call it a horizontal derivation of $ \mathcal{A}$.

(ii) Vertical derivations. Let $\bar {X}$ be a right invariant vector field on $E$. If it satisfies the condition $(\pi_M)_*\bar { X}=0$ it is said to be a vertical vector field. Such vector fields, when lifted to $\Gamma$, are derivations of the algebra $\mathcal{A}$ and are called vertical derivations of \calA .

(iii) As a noncommutative algebra, \calA \ has also inner derivations; they are defined to be
\[
\mathrm{Inn}(\mathcal{A})=\{ad(a):a\in \mathcal{A}\}
\]%
where $(ad(a))(b):=a\ast b-b\ast a$. The mapping $\Phi (a)=ad(a)$, for every $a\in \mathcal{A}$, defines the isomorphism between the algebra $\mathcal{A}$ and the space $\mathrm{Inn}(\mathcal{A})$ as $Z$-moduli.

We call the pair $(\mathcal{A},V)$, where $\mathcal{A}$ is an algebra and $V\subset \mathrm{Der}(\mathcal{A})$ a (sub)module of its derivations, a differential algebra. We shall consider $V = V_1 \oplus V_2 \oplus V_3$ where $V_1, V_2$ and $V_3$ are submoduli of horizontal, vertical and inner derivations of \calA , respectively. Geometry based on the differential algebra $(\calA , V_1 \oplus V_2)$ will be called outer geometry; geometry based on the differential algebra $(\calA , V_3)$ will be called inner geometry. We first summarize the outer geometry.

As the metric $\mathcal{G}:(V_1 \oplus V_2) \times (V_1 \oplus V_2) \rightarrow Z$ for the outer geometry we choose
\[
\mathcal{G}(u,v)=\bar {
g}(u_1,v_1)+\bar {k} (u_2,v_2)
\]
where $u_1,v_{_1}\in V_1,u_2,v_2\in V_2$. The metric $\bar { g}$ is simply the lifting of the
metric $g$ on space-time $M$, and we assume that the metric $ \bar {k}$ is of the Killing type. In \cite{FullJMP} we have demonstrated that, for the case when the group $G$ is semisimple the Killing part of the metric has the form
\[
\bar {k}(\bar{\bar {X}},\bar{\bar {Y}})= \mathcal{B}(\bar X(p), \bar Y(p))
\]
where $\mathcal{B}$ is the Killing form  for the group $G$; it is nondegenerate and is given by
\[
\mathcal{B}(V,W)=\mathrm{Tr}(a d(V)\circ ad(W))
\]
and $V,W$ are elements of the Lie algebra $\mathfrak{g}$ of the group $G$.

It is natural to define the preconnection by the Koszul formula
\[
(\nabla _{u}^{\ast }v)w=\frac{1}{2}[u(\mathcal{G}(v,w))+v(\mathcal{G}%
(u,w))-w(\mathcal{G}(u,v))
\]%
\[
+\mathcal{G}(w,[u,v])+\mathcal{G}(v,[w,u])-\mathcal{G}(u,[v,w]).
\]

It turns out (see \cite[Proposition 3]{FullJMP}) that if $V$ is a $Z$-module of derivations of an algebra $(\mathcal{A},\ast )$ such that $V(Z)=\{0\}$  then, for every symmetric nondegenerate tensor $g:V\times V\rightarrow Z$, there exists exactly one connection $g$-consistent with the preconnection $\nabla ^{\ast }$. For $V_1$ it is the familiar Levy-Civita connection, whereas for $V_2$ it is given by
\[
\nabla _{u}v=\frac{1}{2}[u,v].
\]
For the submodule $V_i,\, i =1,2$, the curvature is defined in the usual way
\[
\stackrel{i}{R}(u,v)w =\stackrel{i}{\nabla}_ u\stackrel{i}{\nabla}_ v w -%
\stackrel{i}{\nabla}_ v\stackrel{i}{\nabla}_ u w -\stackrel{i}{\nabla}_{
[u,v]}w.
\]

For $i=2$, we have
\[
\stackrel{2}{R}(u,v)w =-\frac 14[[u,v],w].
\]

For every endomorphism $T:V_i\rightarrow V_i$, there exists the usual trace $\mathrm{Tr}(T)\in Z$, and we define $\stackrel{i}{R}_{uw}:V_i\rightarrow V_ i$ by
\[
\stackrel{i}{R}_{uw}( v)=\stackrel{i}{R}(u,v) w.
\]
Consequently, we have the Ricci curvature
\[
\stackrel{i}{\mathbf{r} \mathbf{i}\mathbf{c}}(u,w) =\mathrm{Tr}(\stackrel{i}{R}%
_{uw} ),
\]
and the adjoint Ricci operator $\stackrel{i}{\mathcal{R}}: V_i
\rightarrow V_i$
\[
\stackrel{i}{\mathbf{r} \mathbf{i}\mathbf{c}}(u,w) =\stackrel{i}{\mathcal{G}} (%
\stackrel{i}{\mathcal{R}} (u),w)
\]
with $\stackrel{1}{\mathcal{G}} =\bar {g}$ and $\stackrel{2}{\mathcal{G}}=
\bar {k}$. If the metric $\stackrel{i}{\mathcal{G}}$ is nondegenerate, there
exists the unique $\stackrel{i}{\mathcal{R}}$ satisfying the above equation
for every $w\in V_i$.

The curvature scalar is given by
\[
\stackrel{i}{r}=T r(\stackrel{i}{\mathcal{R}} ).
\]

For $V_2$ (for which the usual trace exists) we compute
\[
\stackrel{2}{\mathbf{r} \mathbf{i}\mathbf{c}}(u,w)=\frac 14\bar {k}(u,w)
\]
for every $u,w\in V_ 2$.

The inner geometry was analyzed in \cite{Inner}. Let us only notice that in this case we have a unique connection $\nabla$ associated with the preconnection $\nabla^{*}$, 
\[\nabla_{{\rm a}{\rm d}\,a}{\rm a}{\rm d}\,b=\frac 12[{\rm a}
{\rm d}\,a,{\rm a}{\rm d}\,b],\]
$a, b \in \calA $. Because of the isomorphism between \calA \ and $\mathrm{Inn}\calA $ (as $Z$-moduli and also as Lie algebras) this connection can also be written as 
\[\nabla_ab=\frac 12[a,b] = \frac{1}{2} (a * b - b * a).\]
From the general formula
\[R(a,b)c=\stackrel {}{\nabla}_a\stackrel {}{\nabla}_bc
-\stackrel {}{\nabla}_b\stackrel {}{\nabla}_ac-\stackrel {}{
\nabla}_{[a,b]}c\]
\par
we easily compute
\begin{eqnarray*}
\stackrel {}R(a,b)c&=&-\frac 14[[a,b],c]].\end{eqnarray*}

Unfortunately, it is unclear how to proceed further. Because of the infinite dimensionality of the submodule $V_3$, the usual trace concept is not well defined. However, it turns out that inner derivations are the only one that survive the completion to the von Neumann algebra (see below). This fact, on the one hand, can be interpreted in terms of quantum effects in our model \cite{Conceptual} and, on the other hand, it indicates a certain incompatibility of the standard geometry and quantum mechanics.

\section{Randomization of Geometry}
Let us consider the regular representation of the algebra \calA \ in the Hilbert space $\mathcal{H}^p=L^2(\Gamma^ p)$, for every $p\in E$, $\pi_p:\mathcal{A}\rightarrow \mathcal{B}(\mathcal{H}^p)$, given by
\[
(\pi_p(f)\psi )(\gamma )=\int_{\Gamma^{r(\gamma )}}f(\gamma_1)\psi (\gamma_1^{-1}\circ\gamma )d\gamma_1
\]
where $\mathcal{B}(\mathcal{H}^ p)$ is the algebra of bounded operators in the Hilbert space $\mathcal{H}^p$, and $f\in \mathcal{A} ,\psi\in \mathcal{H}^p, \gamma ,\gamma_1\in\Gamma$. The Haar measure on the group $G$, transferred to each fiber of $ \Gamma $, forms a Haar system on $\Gamma $ \cite[chapter 3]{Paterson}. Let us consider the mapping $\pi : \calA \rightarrow \pi (\calA) = \calM_0$ given by
\[
\pi (a) = (\pi_p(a))_{p \in E}
\]
which is, in fact, an isomorphism of algebras \cite[p. 2506]{Conceptual}. The operator $\pi (a)$ should be understood as an operator in the space $\calH = \int_{\oplus }\calH^p$, the direct integral of the Hilbert spaces $\calH^p$ (\calH \ is isomorphic with the Hilbert space $L^2(E, \calH^p )$ of cross sections of the Hilbert bundle $\{\calH^p\}_{p = E}$). We have demonstrated that every $a\in \mathcal{A}$ generates a random operator $
r_a=(\pi_p(a))_{p\in E}$, acting on a collection of Hilbert spaces $ \{\mathcal{H}^p\}_{p \in E}$ where $\mathcal{H}^p=L^2(\Gamma^p)$ on $\Gamma$ (see \cite{Random})
We recall that an operator $r_{a}$ is a random operator if it satisfies the following conditions \cite[p. 51]{Connes}.

(1) If $\xi_p,\eta_p \in \mathcal{H}^p$ then the function $ E\rightarrow \mathbb{C}$, given by $E\ni p\mapsto (r_a \xi_p,\eta_p)$,$a\in \mathcal{A}$, is measurable (with respect to the usual manifold measure on $E$).

(2) The operator $r_ a$ is bounded with respect to the norm $||r_a||=\,\mathrm{e}\mathrm{s} \mathrm{s}\,\mathrm{s}\mathrm{u} \mathrm{p} ||\pi_p(a)||$.
where ``ess sup'' denotes essential supremum, i.e., supremum modulo zero measure sets.

By a slight abuse of notation let us denote by $\mathcal{M}_{0}$ the algebra of equivalence classes (modulo equality almost everywhere) of bounded random operators $r_{a},a\in \mathcal{A}$ (in analogy with the standard identifying operation in the theory of $L^p$-functions). It can be completed to the von Neumann algebra \calM \ which we call von Neumann algebra of the groupoid $\Gamma $. $\calM_0$ is dense in \calM \ (see Appendix C). It can be shown \cite[p. 2506]{Conceptual} that the mapping $\pi: \calA \rightarrow \calM_0 $ given by
\[
\pi (a) = (\pi_p(a))_{p\in E}
\]
is an isomorphism of algebras. With the help of this isomorphism the differential geometry, as defined by the algebra \calA \ and its derivations, can be transferred to $\calM_0$. In what follows, we study this fact in detail. To do this in a more transparent way, we consider a sheaf of von Neumann algebras rather than a single von Neumann algebra.

\section{A Sheaf of von Neumann Algebras}
Let now, as before, $\Gamma = E \times G$ be the transformation groupoid where, as we remember, $E$ is the total space of the frame bundle over the space-time manifold $M$. The topology on $M$ will be denoted by  $\tau_M$. Let $U \in \tau_M$ and let us consider an open subset $\Gamma_U = \pi_M^{-1}(U) \times G$ of $\Gamma $ where $\pi_M: E \rightarrow M$ is a natural projection.

Let $C^{\infty }_{cf}(\Gamma_U)$ be the algebra of smooth, complex valued functions on $\Gamma_U $ with compact supports along the fibres $\Gamma_p = \pi_E^{-1}(p), \, p \in \pi_M^{-1}(U)$, where $\pi_E: \Gamma \rightarrow E$ is a natural projection. We additionally assume that these functions are bounded. Multiplication in this algebra is defined as convolution in the following way 
\[
(f * g)(\gamma ) = \int_{\Gamma^{r(\gamma )}}f(\gamma_1)g(\gamma_1^{-1} \circ \gamma )d\gamma_1
\]
for $f, g \in C^{\infty }_{cf}(\Gamma_U)$. Actually, compactness along the fibres would be enough to guarantee that the above integral is well defined. 

Let us consider the presheaf $\calF_{cf}$ defined by the mappings $U \mapsto C^{\infty }_{cf}(\Gamma_U)$ and the restriction mappings $\rho_U^V$ given by the usual restriction of a function to a subset of its domain. Let $\calU = (U_j)_{j\in I}$ be an open covering of $M$. We define the family $(f_U)_{U\in \calU }$ where $f_U \in C^{\infty }_{cf}(\Gamma_U)$, such that
\[
f_U|_{\pi_M^{-1}(U)\cap \pi_M^{-1}(V)} = f_V|_{\pi_M^{-1}(U)\cap \pi_M^{-1}(V)}
\]
or, equivalently,
\[
\rho^U_{U \cap V}(f_U) = \rho^V_{U \cap V}(f_V).
\]
This implies that there exists the unique $f \in C^{\infty }_{cf}(\Gamma )$ such that $f|_U = f_U$.

Moreover,  if $f, g \in \calF_{cf}(U)$ then $f*g \in \calF_{cf}(U)$. Therefore, the presheaf $\calF_{cf}$ is a sheaf of noncommutative algebras.

Let us now consider the regular representation $\pi_{0,U}: \calF_{cf}(U) \rightarrow \calB(\calH^U)$, where $\calH^U = \int _{\stackrel{\oplus }{p \in \pi_M^{-1}(U)}}\calH^p $, of the algebra $\calF_{cf}(U)$, given by
\[
\pi_p(f) = f * \psi
\]
for $p \in \pi_M^{-1}(U),\, \psi \in \mathcal{H}^p$.

Let us denote
\[
\calM_0(U) := \{\pi_{0,U}(f): f \in C^{\infty }_{cf}(\Gamma_U) \}.
\]

It should be noticed that $\pi_{0,U}: \calF_{cf}(U) \rightarrow \calM_0(U)$ is an isomorphism of algebras.
In Appendix C it is shown that $\calM_0(U)$ can be completed to the von Neumann algebra $\calM(U)$ as a weak closure $\overline{\calM_0(U)}^{\omega }$. In the same Appendix we prove also that $\calM(U)$ is of the form
\[
\calM(U) = L^{\infty }(U) \bar{\otimes } \calB(L^2(G))
\]
for every $U \in \tau $, with the operations defined in the following way 
\[
(f \otimes A) + (g \otimes A) = (f + g) \otimes A,
\] \[
(f \otimes A) + (f \otimes B) = f \otimes (A+B),
\] \[
(f \otimes A) \circ (g \otimes B) = f \cdot g \otimes A \circ B,
\]
$f, g \in L^{\infty }(U), \, A, B \in \calB (L^2(G))$ for simple elements, and extended (linearly and to the weak closure) for other elements. The functor $U \mapsto \calM (U)$ with the restriction mappings $\rho_V^U: \calM (U) \rightarrow \calM (V)$, given by
\[
\rho_V^U(f \otimes A) = (f|_V) \otimes A,
\]
is a presheaf of von Neumann algebras $\calM (U)$.

By writing down explicitly $\pi_{0,U}$, we see that
\[
\calM_0(U) \cong C^{\infty }_b(U) \otimes \calK (C^{\infty }_c(G \times G))
\]
where $C^{\infty }_b(U)$ is an algebra of smooth bounded functions on $U$, and $\calK (C^{\infty }_c(G \times G))$ is an algebra of integral operators on $L^2(G)$ with kernels belonging to $C^{\infty }_c(G \times G)$. Any operator $A \in \calK (C^{\infty }_c(G \times G))$
is of the form
\[
[A(\psi )](g_1) = \int_G a(g_1, g_2) \psi (g_2) dg_2
\]
for $\psi \in L^2(G), \, g_1, g_2 \in G,\, a \in C^{\infty }_c(G \times G)$. $\mathcal{M}_0(U)$ has a presheaf structure induced by the isomorphism $\pi_{0,U}$; moreover, it is a sub-presheaf of the presheaf of algebras $\mathcal{M}(U)$.

It is clear that the convolution (in the sense of the pair groupoid $G \times G$) corresponds to the composition of integral operators, i.e., we have the isomorphism
\[
\calK (C_c^{\infty } (G \times G), \circ ) \cong (C_c^{\infty }(G \times G), * ).
\]
Therefore, the algebra $\calM_0(U) $ is isomorphic with the algebra $ \widetilde{\calA}(U) = C_b^{\infty }(U) \otimes C_c^{\infty }(G \times G)$, and we can work with either of them.

On the algebra $\calM_0(U)$ we define outer and inner derivations. For the derivation $X \in \mathrm{Der}(C_c^{\infty }(U))$ we define the horizontal derivation $\bar{X} \in \mathrm{Der}_H(\calM_0(U))$ by
\[
\bar{X}(f \otimes A) = (Xf) \otimes A.
\]

On the strength of the above isomorphism of algebras to the derivation $\bar{X}$ there corresponds the derivation $\bar{\bar{X}}$ of the algebra $\widetilde{\calA }(U)$ given by
\[
\bar{\bar{X}}(f \otimes a) = X(f) \otimes a
\]
for $ a \in C_c^{\infty }(G \times G)$. It follows that there is an isomorphism between differential algebras $(\calM_0(U), \mathrm{Der}(\calM_0(U)))$ and $(\widetilde{\calA }(U), \mathrm{Der}(\widetilde{\mathcal{A} }(U)))$.
 
In this way we obtain horizontal derivations of $\calM_0(U)$
\[
\mathrm{Der}_H(\calM_0(U)) = \{\bar{X}: X \in \mathrm{Der}(C^{\infty }(U))\}.
\]We thus have two isomorphisms
\[
\mathrm{Der}(C^{\infty }(U)) \rightarrow \DerH(\calM_0(U)), \;\;\; \mathrm{Der}(C^{\infty }(U)) \rightarrow \DerH(\widetilde{\mathcal{A}} (U))
\]
given by $X \rightarrow \bar{X}$ and $X \rightarrow \bar{\bar{X}}$, respectively.

To deal with vertical derivations let us choose a vector $u \in \mathfrak{g}$, where $\mathfrak{g}$ is the Lie algebra of the group $G$, i.e., $u \in \Der (C^{\infty }(G))$ is a right invariant vector field on $G$. 

We lift $u$ to $\bar{u} \in \Der (C_c^{\infty }(G \times G), *)$. The lifting is understood in the following sense
\[
\bar{u}(a)(g_1,g_2) = [(u_{|1}a)(g_1,g_2) + (u_{|2}a)(g_1, g_2)],
\]
for $a \in C_c^{\infty }(G \times G), g_1, g_2 \in G$, where $u_{|i}, i = 1,2$ are defined as
\[
u_{|1}(a)(g_1,g_2)= (\iota_{g_2})_{*} [ u(a)(g_1) ]
\]
with $\iota_{g_2}: G \rightarrow G \times G$, $\iota_{g_2}(h) = (h, g_2)$, and analogously for $u_{|2}$. 
 
For $a, b \in C_c^{\infty }(G \times G)$ we have
\[
\bar{u}(a*b)(g_1, g_2) = \int_G \bigg( u_{|1}[a(g_1,g_3)b(g_3,g_2)] + u_{|2}[a(g_1,g_3)b(g_3,g_2)] \bigg) dg_3
\] \[
= \int_G [(u_{|1} a)(g_1,g_3) \cdot b(g_3, g_2) + a(g_1,g_3)\cdot (u_{|2}b)(g_3,g_2)]dg_3
\] \[
+ \int_G [(u_{|2} a)(g_1,g_3) \cdot b(g_3, g_2) + a(g_1,g_3)\cdot (u_{|1}b)(g_3,g_2)]dg_3.
\]
The last integral has been added to manifest the Leibniz rule. This could be done since it vanishes. Indeed,
\[
\int_G u[(a \circ \iota_{g_1})(b \circ \iota_{g_2})](g_3)dg_3 
\] \[
= \frac{d}{dt}|_{t=0}\int_G (a \circ \iota_{g_1})(b \circ \iota_{g_2})(\exp (tu) g_3)dg_3
\] \[
= \frac{d}{dt}|_{t=0}\int_G (a \circ \iota_{g_1})(b \circ \iota_{g_2})(g_3)dg_3 =0.
\]

The last line follows from the left invariance of the Haar measure. On the strength of the isomorphism $\mathfrak{g} \rightarrow \Der_V (\calA (U))$ and the isomorphism of algebras $\calA (U) \cong \calM_0(U)$ we have the isomorphism $\mathfrak{g} \rightarrow \Der_V(\calM_0(U))$.

We also have inner derivations
\[
\mathrm{ad}_{f \otimes A}(g \otimes B) = [f \otimes A, g \otimes B]  = (fg) \otimes [A,B].
\]

We now show that $\calM_0(U), \, U \subset M$, determines a sheaf. Indeed, for every $x \in U$ we have the equivalence class of germs, i.e., a stalk $\calM_x$ at $x$. The germs are multiplied in the following way
\[
[(f \otimes A)]_x \circ [(g \otimes B)]_x = [(f \cdot g \otimes A \circ B)]_x.
\]

Let us consider the set of mappings
\[
g : U \rightarrow \bigcup_{x\in U} \calM_x
\]
given by
\[
g(x) = [f \otimes A]_x,
\]
and let us take into account only those mappings of this set which are of the local form, i.e. such that for every $x_0 \in U$ there exist an open neighborhood $W$ of $x_0$ and a $[\xi ]_x \in \calM_x(U)$ with the property $g|_W = [\xi ]_x, \, x \in W$. By standard construction we have the sheaf associated with the presheaf $\calM_0(U)$ (see Appendix A). 

\section{Simple Example}
As a simple example (see \cite{Finite}), let us consider a groupoid $\Gamma_U = U \times G \times G$, $U \subset M$, where $G = \{g_1, \ldots , g_n\}$ is a finite group, with the algebra $\calA (U) = M_{n \times n}(C^{\infty }(U))$ of square matrices with elements from $C^{\infty }(U)$. The center of this algebra is
\[
\calZ(\calA (U)) = \{f\cdot \mathbf{1}_{n\times n}: f \in C^{\infty }(U)\},
\]
For $\gamma_1 = (x,g_1, g'_1), \gamma_2 =(x, g_2, g'_2) \in \Gamma_U$, the composition is $\gamma_1 \circ \gamma_2 = (x, g_1, g_2)$ if $g'_1 = g'_2$ and undefined if otherwise.

Since $ a(x,g_i, g_j) \in \calA(U)$ can be abbreviated to $[a_{ij}] = A(\cdot) \in C_b^{\infty }(U)$, the convolution can be written as
\[
a * b = A(\cdot ) B(\cdot ).
\]

The regular representation of the algebra $\calA(U)$, $\pi_x: \calA \rightarrow \calB(L^2(G))$, is given by
\[
\pi_x(a)\psi = A(x)\psi
\]
for $\psi \in L^2(G) = \C^n, x \in U$, and $\pi = (\pi_x)_{x \in U}$ is a random operator. 
It is clear that in the case of a finite group the formula for $\mathcal{M}_0(U)$ of the previous section reduces to
\[
\calM_0(U) \cong C_b^{\infty }(U) \otimes M_{n \times n}(\C)
\]
The center of the algebra $\calM_0(U)$ is
\[
\calZ(\calM_0(U)) = \{f \otimes \mathbf{1}_{n\times n}: f \in C_b^{\infty }(U)\}.
\]
Because of the above isomorphism the moduli of derivations of the algebra $\calA (U)$ readily ``transfer'' to the moduli of the algebra $\calM_0$.
Let $X \in \Der (C^{\infty }(U))$, and let $\tilde{X}$ be its lifting to $\Der (\calA )$. Then
\[
\tilde{X} ([a_{ij}]) = [Xa_{ij}].
\]
Inner derivations of the algebra $\calA (U)$  are of the form
\[
(\mathrm{ad} \, A)(B) = [A, B].
\]
Taking into account the fact that the weak closure of $C_b^{\infty }(U)$ is $\overline{C_b^{\infty }(U)}^{\omega} = L^{\infty }(U)$, we obtain
\[
\calM(U) = \overline{\cM_0}^{\omega}(U) = L^{\infty }(U) \otimes M_{n \times n}(\C ).
\]
The center of the algebra $\calM(U)$ is
\[
\calZ(\calM (U)) = \{f\otimes \mathbf{1}_{n\times n}: f \in L^{\infty }(U) \}.
\]

Outer derivations of the algebra $\calM_0(U)$ do not survive the completion of  $\calM_0(U)$ to the von Neumann algebra $\calM (U)$, but $\mathrm{ad} \calA$ are still derivations after this completion

\appendix
\section{Appendix: From Presheaves to Sheaves}
Let $M$ be a topological space with topology $\tau $. By the same symbol $\tau $ we denote the category with open sets from this topology as objects and the inclusion mappings $\iota:_U^V: U \hookrightarrow V$ as morphisms. 

\begin{Definition}
A presheaf of algebras on $M$ is a contravariant functor \calF \ from the category $\tau $ to the category {\bf Alg}, i.e., objects $U$ of the catetory $\tau $ go to the objects $\calF (U)$ of the category {\bf Alg} , and the morphisms $\iota_U^V: U \hookrightarrow V$ of the category $\tau $ go to the restriction mappings  $\rho_U^V : \calF(V) \rightarrow \calF(U)$ as morphisms of the category {\bf Alg}.
\end{Definition}

Let $U \subset V \subset W$ then we have
\[
\iota_V^W \circ \iota_U^V = \iota_U^W \Rightarrow \rho_U^W = \rho_U^V \circ \rho_V^W,
\] \[
\iota_U^U = \mathrm{id}_U \Rightarrow \rho_U^U = \mathrm{id}_{\calF(U)}.
\]

\begin{Definition}
Let $\calU = (U_j)_{j \in I}$ be an open covering of a topological space $M$.
A presheaf \calF \ is said to be a sheaf if, for any family $(f_j)_{j \in I}, \, f_j \in \calF (U_j), \, j \in I$, such that 
\[
f_i|_{U_i \cap U_j} = f_j|_{U_i \cap U_j}
\]
for every $i,j \in I$, there exists exactly one (global) element $f \in \calF (\bigcup_{j \in I}U_j)$ such that
\[
f|_{U_j} = f_j, \, j \in I.
\]
\end{Definition}

There exists a canonical method of constructing a sheaf associated with a given presheaf. Let \calF \ be a presheaf on a topological space $(M, \tau )$. For $U \in \tau $, $\calF (U)$ is a set of cross sections of the presheaf \calF \ over $U$. Let $p \in M$, and $f \in \calF (U), g \in \calF (V), \, U,V \in \tau $. We define an equivalence relation in the set $\bigcup_{U \ni p}\calF (U)$ in the following way
\[
f ~\sim_p g \Leftrightarrow \forall_{W \subset U \cap V} f|_W = g|_W.
\]

The equivalence class $[f]_{\sim_p}$ is said to be a germ of the element $f \in \calF (U)$ at the point $p \in M$. This germ is denoted by $\mathbf{f}_p$, and the set of germs at $p$, called the stalk at $p$, is denoted by $\calF_p$. This set inherits the algebraic structure from $\calF (U)$.

Let us denote by $\calF^+(U)$ the set of mappings $\tilde{f}: U \rightarrow \bigcup_{p \in U}\calF_p(U)$ given by
\[
\tilde{f}(q) = \mathbf{f}_q,
\]
$q \in U$, where $f \in \calF (U)$.

We now define the presheaf $\widetilde{\calF}$ by
\[
\widetilde{\calF }(U) = \calF^+(U)_U.
\]
Here the localization (denoted by the subscript $U$) should be understood in the following sense. The mapping $g: U \rightarrow \bigcup_{p \in U} \calF_p(U)$ is a local $\calF^+(U)$-mapping if, for any $q \in U$, there exists an element $\tilde{f} \in \calF^+(U)$ and an open neighbourhood $W \ni q$ such that $g|_W = \tilde{f}|_W$.

It is straghtforward to check that the presheaf $\widetilde{\calF }$ of local $\calF^+(W)$-mappings is a sheaf with the restriction mappings $\rho_U^V: \widetilde{\calF} (V) \rightarrow \widetilde{\calF }(U)$ given by $\rho_U^V(g) = g|_U$.

\section{Appendix: The Center of the Convolution Algebra on a Groupoid}
In this Appendix we show that the center for the algebra \calA \ of smooth, compactly supported functiuons on the transformation groupoid $\Gamma $ is null. In fact, we show this for a more general class of algebras and then specify the result to the case of interest.

Let $X$ be a differential manifold with the Lebesgue measure $\mu $ (let us notice, however, that the proof below remains valid for a locally compact topological space with a Radon measure). Let us consider a pair groupoid $\Gamma = X \times X$ (see \cite{Weinstein}) and the algebra $\calA = C_c^{\infty }(\Gamma , \C)$ with the convolution as multiplication
\[
(a*b)(x,y) = \int_X (a(x,z)b(z,y))d\mu(z)
\]
for $a,b \in \calA $. We shall prove that the center $\calZ (\calA )$ is null, i.e., that $a \in \calZ (\calA )$ implies $a = 0$. The proof will also be valid for the algebra $\calA = C^{\infty }_{cf}(\Gamma, \C )$.

Let us consider the set $K = \mathrm{pr}_2(\mathrm{supp} a)$ where $\mathrm{pr}_2$ is the projection $\mathrm{pr}_2(x,y) = y$. There exists a function $b_1 \in C_c^{\infty }(X, \C)$ such that $\mathrm{supp} b_1 \cap K = \emptyset $. Let us also consider a function $b \in \calA $ of the form $b(z,y) = b_1(z)b_2(y)$, where $b_2 \in C_c^{\infty }(X, \C)$ is given by $b_2(z) = \overline{a(z, y_0)}$ for a fixed $y_0 \in X$. Then
\[
\int_X a(x,z)b(z,y)d\mu (z) = 0
\]
since $a(x,z)b(z,y) = 0$ for any $x,y,z \in X$. On the other hand,
\[
(b*a)(x,y_o) = \int_X b_1(x)b_2(z)a(z,y_0)d\mu (z) 
\] \[
= b_1(x)\int_X b_2(z)a(z,y_0)d\mu (z)
\] \[
b_1(x) \int_x |a(z,y_0)|^2d\mu (z) = 0.
\]
There exists $x \in X$ such that $b_1(x) \neq 0$, therefore
\[
\int_X |a(z,y_0)|^2 d\mu(z) = 0.
\]
Hence $|a(z,y_0)| = 0$, and consequently $a(z,y_0) = 0$ for any $z \in X$. Since $y_0 \in X$ has been chosen arbitrarily, $a = 0$ (this conclusion remains valid also for the algebra $\calA = C_c^{\infty }(X, \C)$ of continuous, compactly suppotrted functions). This ends the proof.

Let us now consider the groupoid $\Gamma = \bigcup_{x \in M} E_x \times E_x$ where $M$ is a space-time, and $E_x$ a fiber over $x \in M$ of the frame bundle $E \rightarrow M$ with the structural group $G$. Let us also consider the algebra $\calA = C_c^{\infty }(\Gamma , \C)$ with the convolution as multiplication
\[
(a*b)(p_1,p_2) = \int_G a(p_1, p_2g)b(p_2g,p_2)d\mu
\]
where $a,b \in \calA , p_1, p_2 \in E, g \in G$.

Let us denote $\calA_x = C_c^{\infty }(\Gamma_x, \C)$ where $\Gamma_x = E_x \times E_x$. The above convolution is an inner operation for the algebra $\calA_x$. Since $\Gamma_x$ is closed in $\Gamma $ then $\calA_x = \calA |\Gamma_x$.
From what we have proved above it follows that $\calZ (\calA_x) = \{ 0 \}$ for any $x \in M$. Of course, if $a \in \calZ (\calA )$ then for every $x \in M$ we have $a|\Gamma_x \in \calZ (\calA_x ) = \{ 0\}$. Therefore, $a = 0$.

Since the pair groupoid $\Gamma = \bigcup_{x \in M} E_x \times E_x$ and the transformation groupoid $\Gamma_1 = E \times G$ are isomorphic (the isomorphism $j: \Gamma_1 \rightarrow \Gamma $ is given by $j(p,g)=(p,pg)$, see also \cite[Proposition 1]{FullJMP}), therefore $\calZ(\calA_1) = \{ 0 \}$ where $\calA_1 = C_c^{\infty }(\Gamma_1, \C)$.

\section{Appendix: The Structure of the Groupoid von Neumann Algebra}
First, let us establish notation. Since we consider the transformation groupoid $\Gamma = E \times G$, and since $E$ is locally trivial it is enough to consider $\Gamma_U = \pi_M^{-1}(U) \times G$, where $U$ is open in $M$. We thus set $\Gamma_U = U \times G \times G$. On this groupoid we have the algebra $\calA (U) = C_{cf}^{\infty }(\Gamma_U ) = \{f \in C^{\infty }(\Gamma_U): (1) \; \forall_{x \in U} \; \mathrm{supp} f_x \; \mathrm{is} \; \mathrm{compact} \; \mathrm{in} \; G \times G, (2) \; \mathrm{sup} |f| < \infty \}$, where $f_x: G \times G \rightarrow \C $ is given by $f_x(g_1, g_2) = f(x, g_1, g_2)$, with the convolution as multiplication
\[
(a_1 * a_2)(x, g_1, g_2) = \int_G a_1(x, g_1, g)a_2(x,g,g_2)dg.
\]

We define the regular representation of the algebra $\mathcal{A}(U)$ on $H = L^2(G)$ by
\[
(\pi_x(a)\psi)(g) = \int_G a(x, g, \bar{g})\psi (\bar{g}) d\bar{g}
\]
for every $x \in U$, and we put $\pi (a) = (\pi_x(a))_{x \in U}$. Now, we introduce the abbreviation $\cM_0(U) = \pi(\calA(U))$, and define the von Neumann algebra $\cM (U)$ as the weak closure of $\cM_0(U) $, $\cM (U) = \overline{\cM_0}^{\omega}(U)$.

\begin{Theorem}
The algebra $\cM (U)$ is isomorphic with the algebra $L^{\infty }(U, \calB (L^2(G)))$. 
\end{Theorem}

We first prove two lemmas.

\begin{Lemma}
The space $C_b^{\infty }(U)$ of smooth bounded functions on $U$ is a wekaly dense subspace of $L^{\infty }(U)$.
\end{Lemma}
 \noindent \textit{Proof.} Let us recall that a sequence $(f_n)$ in $L^{\infty }(U)$ is weakly convergent to $f \in L^{\infty }(U)$ if, for any two elements $\varphi , \psi $ of the Hilbert space $L^2(U)$, one has
\[
((f_n - f) \varphi , \psi )_{L^2(U)} {\rightarrow}_{n \rightarrow \infty } 0. 
\]
As it is well known, every function $f \in L^{\infty }(U)$ is a limit, in the sense of the norm $||\cdot ||_2$ ($||f||_{\infty } = \mathrm{ess}\, \mathrm{sup}_{x \in U}|f(x)|$), of a sequence of simple functions (i.e., of linear combinations of characteristic functions of measurable sets). Of course, convergence in this sense implies the weak convergence. Therefore, it is enough to show that the characteristic function $\chi_A$ of a measurable set $A$ is a weak limit of functions from $C_b^{\infty }(U)$.

It is well known \cite[p. 51]{Brickell} that for any open set $\calO \subset U$ containing $\bar{A}$ there exists a function $\psi \in C^{\infty }(U)$ such that $\psi|\bar{A} = 1$ and $\psi|(U - \calO ) = 0$. Let us remember that the Lebesgue measure is regular. This implies that for the set $\bar{A}$ and every $\epsilon >0$ there exists an open set $\calO_{\epsilon }$ such that $\bar{A} \subset \calO_{\epsilon}$ and $\mu (\calO_{\epsilon } - \bar{A}) < \epsilon $.

Let then $\psi{_\epsilon }$ be a function such that $\psi_{\epsilon }|\bar{A} = 1$ and $\psi_{\epsilon }|(U - \calO_{\epsilon }) = 0$ (as above). It can be easily seen that
\[
((\psi_{\epsilon } - \chi_{A})\xi_1, \xi_2)_{L^2(U)} \rightarrow_{\epsilon \rightarrow  0} 0
\]
for every $\xi_1, \xi_2 \in L^2(U)$. $\Box $

\begin{Lemma}
Space $C_{cf}^{\infty }(U \times G \times G)$ is weakly dense in $L^{\infty }(U, L^2(G \times G))$.
\end{Lemma}

Before we prove this lemma, let us notice that the elements of the space $L^{\infty }(U, L^2(G \times G))$ can be interpreted as operators in the Hilbert space $L^2(U, L^2(G))$ with the scalar product
\[
(\Phi ,\Psi ) = \int_U (\Phi(x), \Psi (x))_{L^2(G)}d\mu (x)
\]
and the operation
\[
(F \cdot \Psi )(x)(g) = \int_G F(x)(g,g_1)\Psi (x)(g_1)dg
\]
for $F \in L^{\infty }(U, L^2(G\times G)),\, \Psi \in L^2(U,L^2(G))$.

A sequence $(F_n)$ in $L^{\infty }(U, L^2(G \times G))$ is weakly convergent to $F \in L^{\infty }(U, L^2(G \times G))$ if, for any two elements $\Phi, \Psi \in L^2(U, L^2(G))$ one has
\[
((F_n - F)\Phi , \Psi )_{L^2(U,L^2(G))} \rightarrow_{n \rightarrow \infty } 0.
\]

\noindent \textit{Proof of Lemma 2.} For $F \in L^{\infty }(U, L^2(G \times G))$ we have the representation
\[
F = \sum_{n=1}^{\infty } \langle F, e_i \rangle _{L^2(G \times G)} e_i
\]
where $\{e_i \}_{i\in \N}$ is an orthonormal basis in the Hilbert space $L^2(G \times G)$. Let $f_i \in \langle F, e_i \rangle _{L^2(G \times G)}, \; i \in \N$. Therefore, $f_i = L^{\infty }(U)$, and from Lemma 1 we have $f_i = \lim_{n \rightarrow \infty } f_i^n$ for a sequence $(f_i^n)$ of functions from $C_b^{\infty }(U)$; this is a weak convergence.

To complete the proof, we will approximate (in the norm $|| \cdot ||_2$) functions $e_i \in L^2(G \times G)$ by functions from $C_c^{\infty }(G \times G)$. It is known that the space $L^2(\R^m)$ contains the space $C_c^{\infty }(\R^m)$ as its dense subspace (with respect to the norm $||\cdot ||_2$) \cite[p. 710]{Maurin}. By applying the smooth decomposition of unity to an open covering of the manifold $G \times G$ by domains of coordinate maps, the above property is transferred from $L^2(\R^m)$ to $L^2(G \times G)$. $\Box $
\vspace{0,2cm}

\noindent
\textit{Proof of Theorem 1.} Let us notice that every function 
$\psi \in L^{\infty }(U, L^2(G \times G))$ (as in Lemma 2) determines the family 
of integral operators $\{\Psi(x)\}_{x \in U}$, $ \Psi (x) \in 
\calB(L^2(G))$ given by 
\[ 
(\Psi(x))(\varphi)(g) = \int_G \psi(x)(g,g_1)\varphi (g_1)dg_1 
\] 
for $\varphi \in L^2(G)$. Let us notice that 
\[ 
\overline{\calK (L^2(G))}^{\omega } = \calB (L^2(G)) 
\] 
where the ``overline $\omega $'' denotes the weak closure of the space of 
integral operators $\calK(L^2(G))$ in the Hilbert space $L^2(G)$.
We should notice that integral operators on a Hilbert space are compact. We also observe that among integral operators there are one-dimensional projectors $P_{\psi } = (\psi_1, \psi )\psi_1$, $\psi_1 \in \calH, ||\psi_1 || = 1$. In fact all operators of finite rank $\cal{F}(H)$ on a Hilbert space $\calH$ are integral operators. 

Now, we make use of the fact that $\overline{\cal{F}(H)}^{\omega } = \calB (\calH)$. Therefore, by $\pi $ we obtain all compact operators. 

Finally, we conclude that $\cM (U) = \overline{\cM_0}^{\omega }(U) = 
L^{\infty }(U, \calB(L^2(G)))$, and obviously $\calM (U)$ is a von Neumann 
Algebra. $\Box $ 

\begin{Theorem}
The algebra $\calM_1 = L^{\infty }(U) \bar{\otimes } \calB (L^2(G))$, where $\bar{\otimes }$ denotes tensor product in the sense of von Neumann algebras (see \cite{Dixmier}), is isomorphic with the algebra $\calM_2 = L^{\infty }(U, \calB(L^2(G)))$. 
\end{Theorem}

\noindent
\textit{Proof.} $\calM_1$ is a von Neumann algebra of operators in the Hilbert space $H_1 = L^2(U) \otimes L^2 (G)$. Any element (of the simple tensor type),  $f \otimes B$, $f \in L^{\infty }(U), \, B \in \calB(L^2(G))$ acts on $\psi \otimes \varphi \in L^2(U) \otimes L^2(G)$ by
\[
(f \otimes B)(\psi \otimes \varphi) = f \psi \otimes B\varphi .
\]
 $\calM_2$ is an operator algebra in the Hilbert space $H_2 = L^2(U, L^2(G))$. Let $A \in \calM_1, \, \Psi \in H_2$, then
 \[
 [A \Psi ](x) = A(x)(f(x)).
 \]
 
 There is an isomorphism $I: H_1 \rightarrow H_2$ given by
 \[
 [I(\psi \otimes \varphi)(x)](\cdot) = \psi(x) \varphi (\cdot),
 \]
 $x \in U$.
 
$\calM_1$ is a weakly closed subalgebra of the algebra $\calB(H_1)$; $\calM_2$ is a weakly closed subalgebra of the algebra $\calB(H_2)$. Of course, $\calB(H_1)$ is isomorphic with $\calB(H_2)$.
 
Let $\{e_i\}$ be a complete orthonormal system of vectors in $L^2(G)$. Elements of the form $f \otimes B, \, f \in L^{\infty }(U), \, B \in \calB(L^2(G))$ generate $\calM_1$; but elements of the form $f \otimes P_{e_i}$, where $P_{e_i}$ is an operator projecting on the vector $e_i$, generate $\calM_1$ as well. We now define, on these generators, a new isomorphism $J: \calM_1 \rightarrow \calM_2$ by
\[
J(f \otimes P_{e_i}) = f \cdot P_{e_i}.
\]
Since the elements of the form $f \cdot P_{e_i}$ generate $\calM_2$, $J$ is an isomorphism of von Neumann algebras. $\Box $


\begin{thebibliography}{cc}
\bibitem{Brickell}
Brickell, F., Clark, R.S. \textit{Differentiable Manifolds. An Introduction}, Van Nostrand, London -- New York, 1970.
\bibitem{Connes}
Connes, A., {\it Noncommutative Geometry}, Academic Press, New York, 1994.
\bibitem{Dixmier}
Dixmier, J., {\it Von Neumann Algebras}, North Holland, Amsterdam, 1981.
\bibitem{Geroch72}
Geroch, R., Einstein Algebras, Commun. Math. Phys. {\bf 26}, 1972, 271-275.
\bibitem{Finite}
Heller, M., Odrzyg\'o\'zd\'z, Z., Pysiak, L., Sasin, W., Noncommutative Unification of General Relativity and Quantum Mechanics. A Finite Model, Gen. Relat. Grav {\bf 36}, 2004, 111-126.
\bibitem{FullJMP}
Heller, M., Pysiak, L., Sasin, W., Noncommutative Unification of General Relativity and Quantum Mechanics, J. Math. Phys. {\bf 46}, 2005, 122501-122516.
\bibitem{Random}
Heller, M., Pysiak, L., Sasin, W., Noncommutative Dynamics of Random Operators, Int. J. Theor. Phys. {\bf 44}, 2005, 619-628.
\bibitem{Inner}
Heller, M., Pysiak, L., Sasin, W., Inner Geometry of Random Operators, Demonstratio Mathematica {\bf 39}, 2006, 971-978.
\bibitem{Conceptual}
Heller, M., Pysiak, L., Sasin, W., Conceptual Unification of Gravity and Quanta, Int. J. Theor. Phys. {\bf 46}, 2007 2494-2512. 
\bibitem{Maurin}
Maurin, K., \textit{Analysis}, Part II, PWN Polish Scientific Publishers, Warszawa and D. Reidel, Dordrecht -- Boston -- London, 1980.
\bibitem{Paterson}
Paterson, A.L.T., {\it Groupoids, Inverse Semigroups and Their Operator Algebras}, Birkh\"auser, Boston -- Basel -- Berlin, 1999.
\bibitem{Weinstein}
de Silva, A., Weinstein, A., {\it Geometric Models for Noncommutative Algebras}, Berkeley Mathematics Lecture Notes, American Mathematical Society, 1999


\end{thebibliography}
\end{document}